\documentclass[
preprint,showpacs,nofootinbib,aps]{revtex4}

\usepackage{float,epsfig}
\usepackage{bm}
\usepackage{graphicx}
\usepackage{subfigure}
\usepackage{dcolumn}
\usepackage{amsmath,amssymb,amsthm}
\def\/{\over}
\begin{document}
\title{\bf Temperature-dependent Casimir-Polder forces on polarizable molecules
}
\author{ Zhiying Zhu $^{1,3}$, Hongwei Yu $^{2,}$\footnote{Corresponding author} and Bin Wang $^1$}
\affiliation{ $^1$ INPAC and Department of Physics, Shanghai Jiao
Tong University, Shanghai 200240, China\\
 $^2 $ Center for Non-linear Science and Department of Physics, Ningbo
University, Ningbo, Zhejiang, 315211, China\\
$^3$ Department of Physics and Electronic Science, Changsha
University of Science and Technology, Changsha, Hunan 410076, China
 }


\begin{abstract}

We demonstrate that the thermal Casimir-Polder forces on molecules near a
conducting surface whose transition wavelengths are comparable
to the molecule-surface separation are dependent on the ambient temperature and  molecular polarization
and they can even be changed from attractive to repulsive via varying
the temperature across a threshold  value for anisotropically
polarizable molecules. Remarkably, this attractive-to-repulsive
transition may be realized at room temperature. Let us note that the predicted repulsion
 is  essentially
a nonequilibrium effect since  the force we calculated on
a ground-state (or an excited-stated) molecule actually contains the contribution
of the absorption (or emission) of thermal photons.


\end{abstract}

\pacs{31.30.jh, 12.20.Ds, 34.35.+a, 42.50.Lc}

\maketitle

 Casimir~\cite{Casimir}  and Casimir-Polder
forces~\cite{CP},
are examples of   striking phenomena that provide convincing
evidence for the reality of  quantum fluctuations of vacuum.  One
usually uses ``Casimir" \cite{Casimir} to describe the force between
two bulk material bodies such as conducting or dielectric plates,
and ``Casimir-Polder" (CP) \cite{CP} to refer to that between a
polarizable object and a material surface. However, the underlying
physical mechanisms are largely the same and there also exist
similar behaviors of the forces. In addition to their fascination in
fundamental research,  Casimir and CP forces are becoming
increasingly important in technological
applications~\cite{Craighead00,Sugimoto07,NT05}.

Recently, the CP forces at finite temperature have attracted a great
deal of attention  on both theoretical and experimental
fronts~\cite{Lifshitz56,Wu00,Antezza04,Antezza05,Harber05,Obrecht07,Mendes07,
Buhmann08-1,Bezerra08,Klimchitskaya08,Pitaevskii08,Derevianko09,zhu09,
Crosse10,Ellingsen10,Ellingsen09,Chwedenczuk10}
(for a recent review, see for example,
Refs.~\cite{Lamoreaux05,Milton09,Bordag09}). The force felt by an
atom near a planar surface at a finite temperature $T$ was first
considered by Lifshitz \cite{Lifshitz56} and is sometimes called the
Lifshitz force. At a distance $z$ which is larger than the thermal
photon wavelength $\hbar c/k_BT$, the attractive Lifshitz force
decays as $1/z^4$ and is proportional to the temperature. In a
previous paper \cite{zhu09}, adopting a approach based upon the
formalism proposed by Dalibard, Dupont-Roc and Cohen-Tannoudji
\cite{Dalibard82,Dalibard84}, which allows a distinct microscopic
treatment to atoms in the ground and excited states in contrast to
the macroscopic approach where atoms are treated as a limiting case
of a dielectric, we have calculated the thermal CP force on a
neutral polarizable two-level atom in interaction with quantized
electromagnetic fields in a thermal bath of temperature $T$ in the
presence of an infinite conducting plane, and analyzed its behavior
in three different regimes of the distance in both the
low-temperature and the high-temperature limits for both the
ground-state and excited-state atoms (see also Ref.~\cite{Mendes07}
for a similar treatment).  Let us note that the same
thermal CP force, was reinvestigated in Ref.~\cite{Ellingsen10}
from the perspective of molecules in
the framework of macroscopic QED \cite{Scheel08}. It is shown there
that the CP force is  independent of the temperature
for a typical molecule placed
near a plane metal surface whose transition wavelength is much
larger than the typical experimental molecule-surface separation in
the nanometer to micrometer range. This is in contrast to the
temperature-dependent CP force for atoms \cite{Lifshitz56,zhu09}. As
a result, the CP forces on molecules with long-wavelength
transitions can not be altered via the ambient temperature.

In this paper,
we demonstrate that
the thermal CP forces on anisotropically polarizable molecules whose transition wavelengths are comparable
to the molecule-surface separation are however temperature-dependent and can even be changed from attractive to repulsive as the
temperature varies across a threshold  value, and therefore can be
dramatically  altered via the ambient temperature.
Let us note here that
the attractive-to-repulsive transitions as a function of temperature
have also been found in the so-called thermodynamic (critical)
Casimir effect~\cite{Bergknoff11,Vasilyev11}.

We model, for simplicity,  a  polarizable molecule as a two-level
system which has stationary states $|-\rangle$ and $|+\rangle$, with
energies $-{1\/2}\hbar\omega_0$ and $+{1\/2}\hbar\omega_0$, and a
level spacing $\hbar\omega_0$  and assume that it is placed at a
distance $z$ from an infinite conducting plane wall.   With a
definition of the molecule's static scalar polarizability
\begin{eqnarray}
\alpha_0=\sum_{j}\alpha_j=\sum_{j~d}{2|\langle
b|\mu_j(0)|d\rangle|^2\/3\omega_{0}\hbar}\;,\label{alpha}
\end{eqnarray}
where  $\alpha_j$ represents polarizability in $j$ direction,
$\mu_j$ is the spatial component of the molecule's dipole moment
and molecule is in its initial state $|b\rangle$,
we have shown that the molecule-wall potentials, which are the
position-dependent corrections to the energy-shifts of the molecule,
are given by~\cite{zhu09},
\begin{eqnarray}
U^{CP}_-=-{3\hbar\omega_{0}\alpha_j\/128\pi\varepsilon_0}
\bigg[{2\/e^{\beta\omega_{0}/c}-1}f_{j}(\omega_{0},z)-g_{j}(\omega_{0},z,\beta)\bigg]\;,\label{CPg}
\end{eqnarray}
for the ground state, and
\begin{eqnarray}
U^{CP}_+={3\hbar\omega_{0}\alpha_j\/128\pi\varepsilon_0}
\bigg[\bigg(2+{2\/e^{\beta\omega_{0}/c}-1}\bigg)f_{j}(\omega_{0},z)
-g_{j}(\omega_{0},z,\beta)\bigg]\;,\label{CPe}
\end{eqnarray}
for the excited state.   Here we have defined
\begin{eqnarray}
f_{x}(\omega_{0},z)=f_{y}(\omega_{0},z)=
{4z^2\omega_{0}^2-c^2\/z^3c^2}\cos(2z\omega_{0}/c)-{2\omega_{0}\/z^2c}\sin(2z\omega_{0}/c)\;,\label{fx}
\end{eqnarray}
\begin{eqnarray}
f_{z}(\omega_{0},z)=-{2\/z^3}\cos(2z\omega_{0}/c)-{4\omega_{0}\/z^2c}\sin(2z\omega_{0}/c)\;,\label{fz}
\end{eqnarray}
\begin{eqnarray}
g_{x}(\omega_{0},z,\beta)=g_{y}(\omega_{0},z,\beta)={64c\/\pi}\sum_{k=-\infty}^\infty\int_0^\infty
du{(uc+k\beta)^2-4z^2\/[(uc+k\beta)^2+4z^2]^3}e^{-\omega_{0}u}\;,\label{gx}
\end{eqnarray}
\begin{eqnarray}
g_{z}(\omega_{0},z,\beta)=-{64c\/\pi}\sum_{k=-\infty}^\infty\int_0^\infty
du{1\/[(uc+k\beta)^2+4z^2]^2}e^{-\omega_{0}u}\;,\label{gz}
\end{eqnarray}
where $\beta={\hbar c/(k_B T)}$ is the wavelength of thermal photons
and summation over repeated indexes is implied. The above result
shows clearly the dependence of the CP potential on polarization of
molecules while isotropy is usually assumed in other
works~\cite{Ellingsen10,Mendes07}.
 The force on the
molecule can be calculated from the potential
\begin{eqnarray}
F^{CP}=-{\partial\/\partial z}U^{CP}\;.\label{force}
\end{eqnarray}

For a given distance $z$,  we can, in the notation of
Ref.~\cite{Ellingsen10}, define a spectroscopic temperature
associated with the molecule transition frequency
$T_{\omega_0}=\hbar\omega_0/k_B$, and a geometric temperature with
the distance  $T_z=\hbar c/zk_B$. For a typical long-wavelength
molecule whose transition wavelength is much larger than the typical
molecule-wall distance, $z\ll c/\omega_0$, $T_{\omega_0}\ll T_z$. We
now analyze how the CP potential behaves as the temperature varies
in three different regimes of temperature, i.e., the low
temperature, where the temperature $T$ is much lower than the
spectroscopic temperature ($T\ll T_{\omega_0}\ll T_z$), the
intermediate temperature, where the temperature $T$ is much higher
than the spectroscopic temperature but much lower than the geometric
temperature ($T_{\omega_0}\ll T\ll T_z$), and the high temperature,
where the temperature $T$ is much higher than the geometric
temperature ($T_{\omega_0}\ll T_z\ll T$).

Let us start with a typical molecule with long-wavelength transitions such that
$z\ll c/\omega_0$. Now the oscillating functions, Eqs.~(\ref{fx}) and
(\ref{fz}), can be written as
\begin{eqnarray}
f_{x}(\omega_{0},z)=f_{y}(\omega_{0},z)\approx -{1\/z^3}+{2\omega_0^2\/zc^2}-{6\omega_0^4z\/c^4}\;,\label{fx1}
\end{eqnarray}
\begin{eqnarray}
f_{z}(\omega_{0},z)\approx -{2\/z^3}-{4\omega_0^2\/zc^2}+{4\omega_0^4z\/c^4}\;.\label{fz1}
\end{eqnarray}
In the low temperature regime, i.e., when $T\ll T_{\omega_0}\ll
T_z$, we have $z\ll c/\omega_0\ll\beta$. So the exponential
function, ${2\/e^{\beta\omega_{0}/c}-1}$, approaches zero. And
functions $g_{i}(\omega_{0},z,\beta)$ can be approximated as
\begin{eqnarray}
g_{x}(\omega_{0},z,\beta)=g_{y}(\omega_{0},z,\beta)={1\/2}g_{z}(\omega_{0},z,\beta)\approx-{1\/z^3}\;.\label{g1}
\end{eqnarray}
Plugging Eqs.~(\ref{fx1})-(\ref{g1}) into Eqs.~(\ref{CPg}) and
(\ref{CPe}) and taking derivative of the potentials with
respect to the distance $z$, we obtain the forces acting on the
molecules in both ground  and excited states,
\begin{eqnarray}
F^{CP}_-\approx F^{CP}_+\approx-{\hbar\/4\pi\varepsilon_0}{9\omega_0\/32z^4}(\alpha_\parallel+2\alpha_z)\;,\label{regime1}
\end{eqnarray}
where $\alpha_\|=\alpha_x+\alpha_y$ denotes the molecular polarizability
in the transverse direction. This is just the usual
temperature-independent van der Waals force.

If the temperature is higher than the spectroscopic temperature,
$T\gg T_{\omega_0}$, we have $\beta\ll c/\omega_0$. In this case the
exponential function, ${2\/e^{\beta\omega_{0}/c}-1}$ can be
approximated by ${2c\/\beta\omega_0}-1$, and we find
\begin{eqnarray}
g_{x}(\omega_{0},z,\beta)=g_{y}(\omega_{0},z,\beta)={1\/2}g_{z}(\omega_{0},z,\beta)\approx-{2c\/\beta\omega_0z^3}\;.\label{g2}
\end{eqnarray}
Consequently, in the spectroscopic high-temperature limit $T\gg T_{\omega_0}$, the forces become
\begin{eqnarray}
F^{CP}_-\approx-{\hbar\/4\pi\varepsilon_0}\bigg[{9\omega_0\/32z^4}(\alpha_\parallel+2\alpha_z)
+{3\omega_0^2\/8c\beta z^2}(\alpha_\parallel-2\alpha_z)
+{9{\omega_0}^4\/8c^3\beta}\bigg(\alpha_\parallel-{2\/3}\alpha_z\bigg)\bigg]\;,\label{cp2-}
\end{eqnarray}
\begin{eqnarray}
F^{CP}_+\approx-{\hbar\/4\pi\varepsilon_0}\bigg[{9\omega_0\/32z^4}(\alpha_\parallel+2\alpha_z)
-{3\omega_0^2\/8c\beta z^2}(\alpha_\parallel-2\alpha_z)
-{9{\omega_0}^4\/8c^3\beta}\bigg(\alpha_\parallel-{2\/3}\alpha_z\bigg)\bigg]\;. \label{cp2+}
\end{eqnarray}
Let us now further analyze these forces in two subcases, i.e.,  $T_{\omega_0}\ll T\ll T_z$ and  $T_{\omega_0}\ll T_z\ll T$.
In the intermediate temperature regime, where the temperature $T$ is
much higher than the spectroscopic temperature but much lower than
the geometric one ($T_{\omega_0}\ll T\ll T_z$), we have $z\ll
\beta\ll c/\omega_0$. So, the first term  in both Eqs.~(\ref{cp2-})
and (\ref{cp2+}) is much larger than the other terms. As a result,
the forces are still independent of the temperature in the leading
term. However, if the temperature moves to the high temperature
regime, i.e., when $T_{\omega_0}\ll T_z\ll T$, or equivalently
$\beta\ll z\ll c/\omega_0$, then the second term in Eqs.~(\ref{cp2-})
and (\ref{cp2+}) will be dominant
when $\beta<|{\alpha_\parallel-2\alpha_z\/\alpha_\parallel+2\alpha_z}|
{4\omega_0z^2\/3c}$, or equivalently
$T>|{\alpha_\parallel+2\alpha_z\/\alpha_\parallel-2\alpha_z}|{3\hbar
c^2\/4k_B\omega_0z^2}$, for molecules which are not isotropically polarized.
In other words, above a threshold  temperature,
$T_{threshold}\sim
|{\alpha_\parallel+2\alpha_z\/\alpha_\parallel-2\alpha_z}|{3\hbar
c^2\/4k_B\omega_0z^2}$, the force becomes temperature-dependent. If
molecules are isotropically polarized, the second term in both
Eqs.~(\ref{cp2-}) and (\ref{cp2+}) vanishes. In this case, the
threshold  temperature appears at $T_{threshold}\sim {3\hbar
c^4\/4k_B\omega_0^3z^4}$ above which the third term dominates so that
the CP force varies with temperature.
Noteworthily, for a molecule in its
ground state which is anisotropically polarized such that
$\alpha_\parallel-2\alpha_z<0$, the CP force changes sign at the
threshold  temperature and turns to repulsive from attractive once the
point is crossed, whereas for a molecule in its excited state, the
force change sign at the threshold  temperature when
$\alpha_\parallel-2\alpha_z>0$. However,  for isotropically
polarizable molecules, the force is always attractive for the ground
state, but it can change sign and become repulsive for excited
states. So the properties of the CP force depend crucially on the
polarization  of molecules.

Let us now estimate the threshold  temperature for a typical
long-wavelength molecule. Taking LiH whose vibrational transition
frequency is $\omega_0=4.21\times10^{13}$ Hz \cite{Buhmann08} as an
example, we find that at a distance of $z=1\,\mu$m
($z\omega_0/c=0.14$), the threshold  temperature for a ground state
anisotropic  molecule with only longitudinal polarization is
$T_{threshold}\sim 1.2\times10^4$ K, and that for an excited state
molecule which is isotropically polarizable is $T_{threshold}\sim
6.4\times10^5$ K. These threshold  temperatures
are not currently accessible
in experiments.
As a result, the temperature-dependent terms in Eqs.~(\ref{cp2-}) and
(\ref{cp2+}) which come from the
oscillating functions $f_{j}(\omega_{0},z)$ can be
ignored in practical experimental sense. So, for typical long-wavelength molecules,
 the thermal CP forces are essentially temperature-independent~\cite{Ellingsen10}, at least within the  experimentally
accessible temperature regimes.
However, this may change dramatically
when the molecule-wall distance is comparable to the transition
wavelength of the molecule, as we will demonstrate next by numerical
analysis.

\begin{figure}[htbp]\centering
{\subfigure[]{\label{fig1a}
\includegraphics[height=1.5in,width=2.0in]{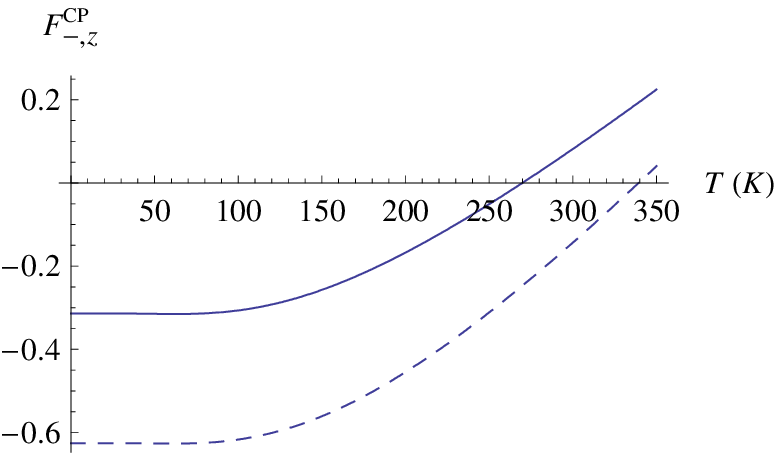}}
\subfigure[]{\label{fig1b}
\includegraphics[height=1.5in,width=2.0in]{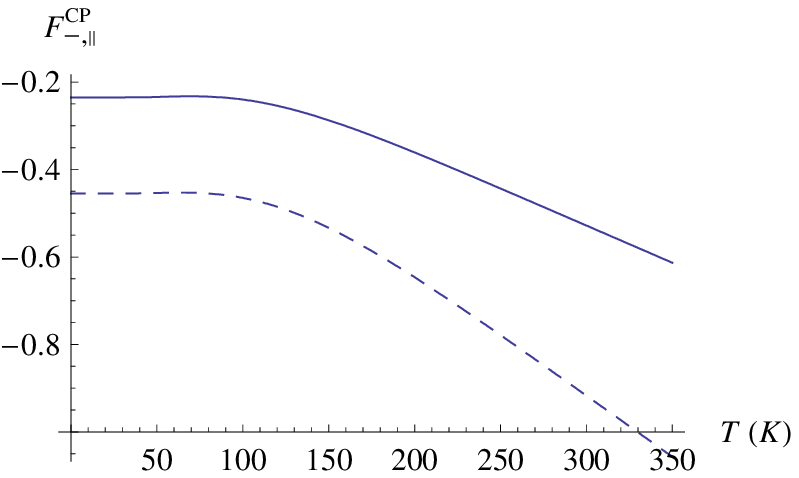}}
\subfigure[]{\label{fig1c}
\includegraphics[height=1.5in,width=2.0in]{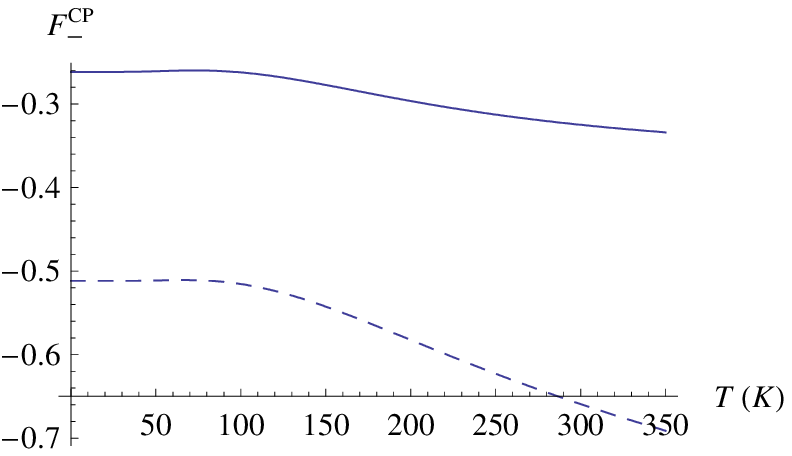}}}
\caption{The temperature-dependent CP force between a ground
state LiH molecule and a conducting plane wall, when the molecule
is polarizable (a) in the $z$-direction, (b) in the direction
parallel to the conducting plane wall,
(c) isotropically. Here the distance is
respectively $z\sim c/\omega_0\simeq7\,\mu$m (solid line) and
$z=6\,\mu$m (dashed line). The force is in the unit of ${\hbar c
\omega_0^2\alpha_0/(128\pi\varepsilon_0)}.$}\label{fig1}
\end{figure}

Taking LiH as an example again, we have plotted, in
Fig.~\ref{fig1}, the CP force as a function of temperature $T$
for a LiH molecule in its ground state at a distance $z\sim
c/\omega_0\simeq 7\,\mu$m and $z=6\,\mu$m respectively.
The plots show clearly the temperature dependence of the force.
One can see, from Fig.~\ref{fig1a}, that for a ground
state LiH molecule which is polarizable only in the $z$-direction,
the CP force becomes
positive when the temperature is above a threshold  value. For the
molecule-wall distance $z=7\,\mu$m, numerical computation reveals
that the threshold  temperature occurs at about only $270$ K. But for
the molecules which are polarizable only in the direction parallel to the
conducting plane wall or polarizable isotropically, the temperature
dependent CP forces are always attractive as shown in
Figs.~\ref{fig1b} and \ref{fig1c}.
This means that a repulsive CP force may be observed at room
temperature for a longitudinally polarizable molecule in its ground
state and a manipulation of the CP
 forces via the ambient temperature  can be demonstrated in laboratory.
If we decrease the molecule-wall distance, for instance, if we place
the LiH molecule at a distance $z=6\,\mu$m, the CP force as a
function of temperature becomes what is plotted with dashed lines in
Fig.~\ref{fig1}. Now the threshold  temperature appears at $340$ K.
So, for a given molecule, the threshold  temperature increases with the decrease of
the distance and it goes up too high to reach in experiment for
$z\ll c/\omega_0$.

\begin{figure}[htbp]
\centering
\includegraphics[width=3.0in]{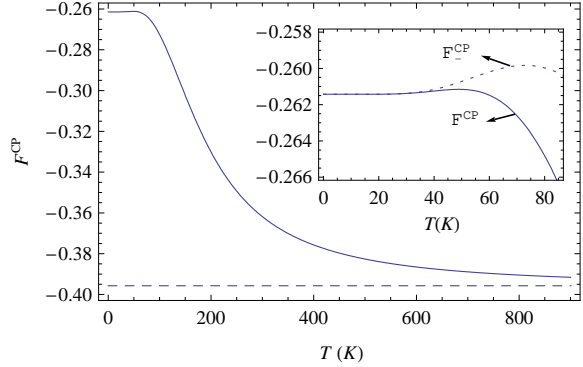}
\caption{ The behavior of the average CP force for an isotropically
polarizable LiH molecule as a function of temperature $T$ for a
distance $z\sim c/\omega_0(\simeq7\,\mu$m). Here the force is in the
unit of ${\hbar c\omega_0^2\alpha_0/(128\pi\varepsilon_0)}$. When
the temperature $T<30$ K, the average force is essentially equal to
that of a ground state molecule (dotted line). In the high
temperature regime $T_z\sim T_{\omega_0}\ll T$, the force
approximates to a temperature-independent value (dashed
line).}\label{fig4}
\end{figure}

In the above discussions, we have examined the CP force for a
molecule in its ground and excited states. Now we turn our attention
to the thermal average of the force for a molecule in
equilibrium with thermal photons, which can be written as
\begin{eqnarray}
F^{CP}&=&{1\/1+e^{-\omega_{0}\beta/c}}F^{CP}_- +\bigg(1-{1\/1+e^{-\omega_{0}\beta/c}}\bigg)F^{CP}_+
\;.\label{average}
\end{eqnarray}
First, we consider molecules whose transition wavelengths are larger
than the molecule-wall distance, $z\ll c/\omega_0$.
In the low temperature regimes,
$T\ll T_{\omega_0}$ ($\omega_{0}\beta/c\gg1$),  the average
force is essentially given by the force of a ground state
molecule, i.e., Eq.~(\ref{regime1}), which is independent of
temperature. In the intermediate and high temperature regimes, the
temperature is much higher than the spectroscopic temperature, i.e.,
$T_{\omega_0}\ll T$ ( $\omega_{0}\beta/c\ll1$). So,  the
contributions of the ground state and the excited one are almost
equally weighted in the thermal average and therefore the
temperature-dependent parts (refer to Eqs.~(\ref{cp2-}) and
(\ref{cp2+})) cancel, leading to  a temperature-independent final
result the same as  Eq.~(\ref{regime1}).  Thus, the average
force is temperature-independent over the entire range. This is
in coincidence  with the result in Ref.~\cite{Ellingsen10}. However,
what we want to further show here is that this temperature
independence of the average force is not universal. As a matter
of fact, when the molecule-wall distance is comparable to the
transition wavelength of the molecule, i.e., $z\sim c/\omega_0$, the
average force becomes temperature-dependent in some regime, as
is demonstrated in Fig.~\ref{fig4} where the average force of a
isotropically polarizable LiH molecule at distance
$z\sim c/\omega_0\simeq7\,\mu$m is plotted.
The figure shows that when the temperature $T< 30$ K, the
average force is equal to that for a ground state molecule. This
comes as no surprise, since  the  energy of thermal photons in this
case is much smaller than the transition energy $\hbar\omega_0$ so
that transitions from the ground state to  excited states are
virtually impossible. However, as the temperature goes higher,  the
average force decreases obviously with the increase of
temperature until temperature reaches the high temperature regime
$T_z\sim T_{\omega_0}\ll T$ where the average force becomes
temperature-independent again. This temperature-independence of the
average force in the high temperature regime is actually
expected on the ground of analytical analysis, since a combination
of
 Eq.~(\ref{CPg}), (\ref{CPe}), (\ref{force}), (\ref{g2}) and (\ref{average}) leads
 to the following average force
\begin{eqnarray}
F^{CP}=
-{3\hbar\omega_{0}\alpha_j\/128\pi\varepsilon_0}~{1-e^{-\omega_{0}\beta/c}\/1+e^{-\omega_{0}\beta/c}}~
{\partial\/\partial z}g_j(\omega_{0},z,\beta)
\approx-{\hbar\/4\pi\varepsilon_0}{9\omega_0\/32z^4}(\alpha_\parallel+2\alpha_z)\;,\label{average2}
\end{eqnarray}
which is temperature-independent and coincides with the average force for a typical long-wavelength molecule.

Now it is worth pointing out that our investigation of the thermal CP force is based on a 
Quantum Field Theory
treatment of the molecule-field interaction. The force we calculated on
a ground-state (or an excited-stated) molecule actually contains the contribution
of the absorption (or emission) of thermal photons which is essentially
a nonequilibrium effect, and it does not always agree with what is obtained from a macroscopic
calculation in Lifshitz theory, where an atom(or a molecule)
is treated as a rarefied medium in thermal equilibrium with a dielectric surface.
Let us note that details of whether and under what conditions Lifshitz theory may be
used to describe thermal CP forces on atoms or molecules have been discussed
in Ref.~\cite{Buhmann08-1} in the framework of QED.
It is also worth pointing out that  repulsive nonequilibrium Casimir forces have recently
been studied in Refs.~\cite{Bimonte11,Golyk12}.

To summarize, we have analyzed the thermal CP forces for  neutral
polarizable molecules near  an infinite conducting plane wall
and we find that the CP forces on molecules with transition
wavelengths which are long as compared to the molecule-wall
distance is dependent on the temperature,
but this temperature dependence can
not be exploited to alter the CP force via ambient temperature
since the threshold  temperature beyond which the CP forces become
temperature-dependent is not currently accessible in experiments.
However, if the molecule-wall separation is comparable to the
transition wavelength, then CP forces on molecules display
interesting temperature dependence at room temperature which allows
us to dramatically manipulate them  via the ambient temperature,
and the force can even be changed from attractive to repulsive
in experiments at room temperature.

\begin{acknowledgments}
This work was supported in part by the National Natural Science
Foundation of China under Grants  No. 11075083, 10935013, 11005013 and No.
11105020; the Zhejiang Provincial Natural Science Foundation of
China under Grant No. Z6100077;  the National Basic Research Program
of China under Grant No. 2010CB832803; the PCSIRT under Grant No.
IRT0964;  the China Postdoctoral Science Foundation under Grant No.
2011M500764 and No. 2012T50414; the Research Foundation of Hunan Provincial 
Education Department under Grant No. 10C0377 and No. YB2011B038, 
and the K. C. Wong Magna Fund in Ningbo University.
\end{acknowledgments}

\end{document}